\documentclass[twocolumn,showpacs,amsmath,amssymb,prd]{revtex4-1}
             
\usepackage{graphicx}
\usepackage{epsfig}
\usepackage{dcolumn}
\usepackage{bm}
\usepackage{mathrsfs,amsmath}

\def\beq{\begin{equation}}
\def\enq{\end{equation}}  
\def\ba{\begin{eqnarray}}
\def\ea{\end{eqnarray}}
\def\<{\langle}
\def\>{\rangle}

\begin{document}

\title{Future constraints on the Hu-Sawicki modified gravity scenario} \author{Matteo Martinelli$^1$, Alessandro Melchiorri$^1$, Olga Mena$^2$, Valentina Salvatelli$^1$ and Zahara Giron\'es$^2$.} 
 \affiliation{$^1$Physics Department and INFN, Universita' di Roma ``La Sapienza'', Ple Aldo Moro 2, 00185, Rome, Italy.}
 \affiliation{$^2$IFIC-CSIC and Universidad de Valencia, Valencia, Spain}

\begin{abstract}

We present current and future constraints on the Hu and Sawicki modified gravity scenario. This model can reproduce a late time accelerated universe and evade solar system constraints. While current cosmological data still allows for distinctive deviations from the cosmological constant picture, future measurements of the growth of structure combined with Supernova Ia luminosity distance data will greatly improve present constraints. 

\end{abstract}


\date{\today}
\maketitle

\section{Introduction}
 Astrophysical observations have led to the inference that our
 universe is approximately  flat and its mass-energy budget
 consists of $5\%$ ordinary matter, $22\%$ non-baryonic 
dark matter, plus a dominant negative-pressure component that 
accelerates the Hubble expansion~\cite{Komatsu:2010fb,Percival:2009xn,Reid:2009xm,Amanullah:2010vv}. The current accelerated 
expansion can be explained as the presence
 of a Cosmological Constant (CC hereafter) associated to the energy of the vacuum in Einstein's equations. However, the naively theoretically expected value exceeds the measured one 
by $123$ orders of magnitude 
and it needs to be cancelled by extreme fine-tuning. A dynamical alternative attributes the accelerated expansion to a 
cosmic scalar field,
\emph{quintessence}~\cite{Caldwell:1997ii,Zlatev:1998tr,Wang:1999fa,Wetterich:1994bg,Peebles:1987ek,Ratra:1987rm} in which 
case the equation of state $w$ could vary 
over time. However, these quintessence models are not better than the CC scenario as regards fine-tuning.

In this paper we focus on the third possible scenario, in which the gravitational sector is modified, as an alternative to 
explain the observed cosmic acceleration (see Refs.~\cite{Clifton:2011jh,Tsujikawa:2010zza,DeFelice:2010aj} and references therein).  
Although this requires the modification of Einstein's equations of
gravity on very large distances~\cite{Dvali:2000hr}, or on small curvatures~\cite{Carroll:2003wy,Capozziello:2003tk,
Vollick:2003aw}, this is not unexpected for an 
effective 4-dimensional description of higher dimensional
theories. \emph{Modifications of gravity} have been examined 
in the context of accelerated expansion. The proposed modified gravity
models have extra spatial dimensions or an action which is non
linear in the curvature scalar, that is, these models include
extensions of the Einstein-Hilbert action, for instance, 
to higher derivative theories~\cite{Carroll:2004de}, 
scalar-tensor theories or generalized functions of the Ricci 
scalar $f(R)$. Modified gravity models have been confronted with current and future data extensively in the literature~\cite{Mena:2005ta,Zhang:2005vt,Ishak:2005zs,Amarzguioui:2005zq,Tang:2006tk,Huterer:2006mva,Borowiec:2006hk,DeFelice:2007ez,Jain:2007yk,Guzzo:2008ac,Zhao:2008bn,Girones:2009nc,Giannantonio:2009gi,Zhao:2010dz,Moldenhauer:2010zz,Lombriser:2010mp,Zhao:2011te}. However, it is well known that $f(R)$ gravity models that produced late time acceleration also have problems to pass solar system tests~\cite{Chiba:2003ir,Navarro:2005gh,Olmo:2005jd,Olmo:2005zr,Capozziello:2005bu,Navarro:2006mw,Hu:2007nk,Olmo:2006eh,Amendola:2007nt,Capozziello:2007ms,Tsujikawa:2008uc}.
The reason is that $f(R)$ gravity theories introduce a scalar degree of freedom given by $f_R\equiv df/dR$ that, for the background cosmological density, 
is very light. As a consequence, it produces a long-range fifth force, 
leading to a dissociation of the space-time curvature from the
local density. Then, the metric around the sun is predicted to be different than what is observed. Chiba 
\cite{chiba:2006jp} has shown the conditions under which a given
$f(R)$ model is equivalent to a scalar-tensor theory with Parametrized Post-Newtonian (PPN) parameter 
$\gamma=1/2$, far outside the range allowed
by observations, $|\gamma -1| < 2.3 \cdot 10^{-5}$~\cite{Will:2005va}. However, some $f(R)$ theories are still viable: 
the scalar field mass could be
large and therefore it would not have an effect at solar system scales.
Another possibility is a scale dependent scalar field mass, as in
the chameleon mechanism~\cite{Khoury:2003rn,Cembranos:2005fi,Faulkner:2006ub,Capozziello:2007eu,Brax:2008hh}. 
In chameleon cosmologies, 
the effective mass of the scalar field becomes very large in 
high density environments (as in the Sun's interior) and the
induced fifth force range would be below the detectability level of
gravitational experiments. 
 
Among a plethora of $f(R)$ models, we focus here on the one proposed by Hu
and Sawicki~\cite{Hu:2007nk} (HS hereafter). This model is able to reproduce the late time accelerated universe, but with 
distinctive deviations from a cosmological
constant. The model has also been shown to satisfy the conditions needed to produce a cosmologically viable expansion 
\cite{Martinelli:2009ek}. More interestingly, 
the model is designed to posses a chameleon mechanism that allows to easily evade solar system constraints. The authors of 
Ref.~\cite{Martinelli:2009ek} have analyzed the HS model exploiting current cosmological data. Here we update these results using the most recent cosmological data and we present the expected constraints on the HS model from future Baryon Acoustic Oscillation (BAO) surveys and future Supernovae Ia (SNIa) luminosity distance measurements. 

The structure of the paper is as follows. In Sec.~\ref{sec:i} we briefly describe the HS model. In Sec.~\ref{sec:ii} we present constraints on this model using the most recent cosmological data. Section \ref{sec:iii} describes the method used here to forecast data from future BAO and SNIa surveys. Future constraints from the former data are presented in Sec.\ref{sec:iv}. We draw our conclusions in Sec.~\ref{sec:v}.

\section{The model} \label{sec:i}

 We briefly review below the basic equations and results of the HS model. This model has a modified Einstein-Hilbert action:

\begin{equation}\label{eq:action}
S=\int{d^4x \sqrt{-g} \big[\frac{R+f(R)}{2\kappa^2}+{\cal L}_{m}\big]}~,
\end{equation}

\noindent where ${\cal L}_m$ is the matter lagrangian, $\kappa^2=8\pi G$ and

\begin{equation}\label{eq:fr}
f(R)=-m^2\frac{c_1\big(\frac{R}{m^2}\big)^n}{1+c_2\big(\frac{R}{m^2}\big)^n}~,
\end{equation}

\noindent with $m^2=\kappa^2\rho_0/3$, being $\rho_0$ the average density today and $c_1$, $c_2$ and $n$ as free parameters.

Varying the action Eq.~(\ref{eq:action}) with respect to the metric $g^{\mu\nu}$ one obtains the modified Einstein equations
\begin{equation}
G_{\mu\nu}+f_RR_{\mu\nu}-\big(\frac{f}{2}-\Box f_R\big)g_{\mu\nu}-\nabla_\mu
\nabla_\nu f_R=k^2T_{\mu\nu}~,
\end{equation}
\noindent where $f_R=df/dR$ and $f_{RR}=d^2f/dR^2$. Assuming a flat Friedmann Robertson Walker (FRW) metric, the modified Friedmann equation reads
\begin{equation}
H^2-f_R(HH'+H^2)+\frac{f}{6}+H^2f_{RR}R'=\frac{\kappa^2\rho}{3}~,
\end{equation}
\noindent with $'\equiv d/dlna$. Defining the new variables $y_H=(H^2/m^2)-a^{-3}$ and $y_R=(R/m^2)-3a^{-3}$ the Friedmann 
equation can be written as a system of two 
ordinary differential equations:
\begin{equation}
y'_H=\frac{y_R}{3}-4y_H~;
\end{equation}

$$y'_R=9a^{-3}-\frac{1}{y_H+a^{-3}}\frac{1}{m^2f_{RR}}$$
\begin{equation}
\times\big[y_H-f_R\big(\frac{y_R}{6}-y_H-\frac{a^{-3}}{2}\big)+\frac{f}{6m^2}\big]~.
\end{equation}

In order to compare the HS model with the cosmological constraints
usually derived under the assumption of a dark energy fluid, it is useful
to introduce an {\it effective} dark energy component with a present energy density $\tilde{\Omega}_x=1-\tilde{\Omega}_m$ and 
a equation of state $w$, where
$\tilde{\Omega}_m$ is the effective matter energy density at present time. 
Of course, in reality, no dark energy component is present and the only
component of the universe is matter, being modified gravity the responsible for  the accelerated expansion. Considering the 
Friedmann equation

\begin{equation}
\frac{H^2}{H_0^2}=\frac{\tilde{\Omega}_m}{a^3}+\tilde{\Omega}_x
e^{\int^1_a{da\frac{3[1+w(a)]}{a}}}~,
\end{equation}

\noindent the effective equation of state parameter $w$
for the dark energy component is given by

\begin{equation}
w=-1-\frac{1}{3}\frac{y'_H}{y_H}~.
\end{equation}

The free parameters $c_1$ and $c_2$  that appear in Eq.~(\ref{eq:fr}) can be expressed as a function of the effective 
density parameters

\begin{equation}
\frac{c_1}{c_2}\approx6\frac{\tilde{\Omega}_x}{\tilde{\Omega}_m}~;
\end{equation}
and
\begin{equation}
\frac{c_1}{c_2^2}=-\frac{f_{R_0}}{n}\big(\frac{12}{\tilde{\Omega}_m}-9\big)^{n+1}~.
\end{equation}

\begin{figure}[h!]
\centering
\includegraphics[width=8cm]{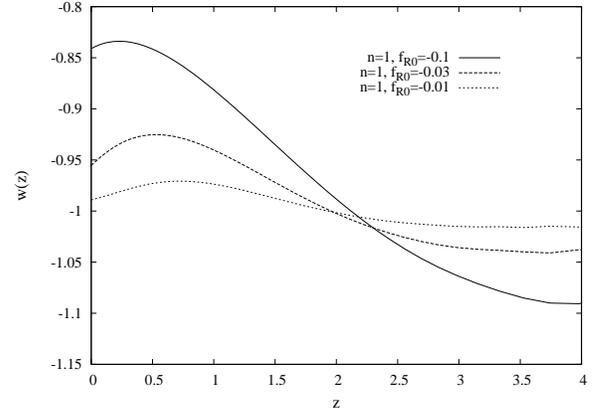}
\caption{The equation of state parameter $w$ for $n=1$ and $f_{R_0}=-0.1, -0.03$ and $-0.01$ (solid, dashed and dotted 
lines). $\tilde{\Omega}_m$
is set to $0.24$.}
\label{wn1}
\end{figure}

\begin{figure}[!h]
\centering
\includegraphics[width=8cm]{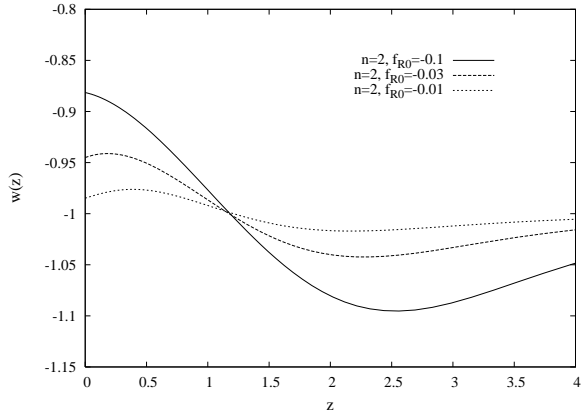}
\caption{The equation of state parameter $w$ for $n=2$ and $f_{R_0}$ equal
to $-0.1$ (solid line), $-0.03$ (dashed line) and $-0.01$ (dotted line). $\tilde{\Omega}_m$ is set to $0.24$.}
\label{wn2}
\end{figure}

Using these last two equations we can relate $c_1$ and $c_2$ to the free parameters of the model, $n$ and $f_{R_0}\equiv 
f_R (lna =0)$, and to $\tilde{\Omega}_m$. 
The parameter $f_{R_0}$ is constrained to $|f_{R_0}|\lesssim 0.1$ by solar system  tests \cite{Hu:2007nk} and therefore we will not 
investigate larger values in the next 
sections. Figures \ref{wn1} and \ref{wn2} show the behavior of $w$ as a function of the redshift $z$ for different values 
of $f_{R_0}$ and $n$.

As we can see in Fig.\ref{wn1} and Fig.\ref{wn2} the equation of state parameter $w$ follows a peculiar behavior as a 
function of the redshift.
At the present time ($z=0$) $w$ has always a value higher than the
one predicted by the $\Lambda$CDM model
($w=-1$) and, moving towards higher redshifts, it decreases towards the phantom region, i.e., taking values lower than $-1$. For
even higher redshifts, $w$ moves asymptotically towards $-1$. If the absolute value of $f_{R_0}$ is decreased, $w$ gets closer to $-1$, while if the 
parameter $n$ is increased the phantom crossing occurs at lower redshifts.

After solving the background equations in terms of the new variables $y_H=(H^2/m^2)-a^{-3}$ and $y_R=(R/m^2)-3a^{-3}$ we 
are able to predict the expected
theoretical values for a set of observables in order to constrain the free 
parameters of the HS model ($n$ and $f_{R_0}$). We consider here the luminosity distance
\begin{equation}\label{eq:dl}
d_L(a)=\frac{1}{a}\int_a^1{\frac{da}{a^2H(a)}}=\frac{1}{aH_0}\int_a^1
{\frac{da}{a^2\sqrt{\tilde{\Omega}_m(y_H+a^{-3})}}}~,
\end{equation}

\noindent as well as the Hubble parameter

\begin{equation}\label{eq:H}
H(a)=\sqrt{\tilde{\Omega}_mH_0^2(y_H+a^{-3})}~,
\end{equation}

\noindent and the angular diameter distance
\begin{equation}\label{eq:da}
d_A(a)=\int_a^1{\frac{da}{a^2H(a)}}=\frac{1}{H_0}\int_a^1
{\frac{da}{a^2\sqrt{\tilde{\Omega}_m(y_H+a^{-3})}}}~.
\end{equation}

We also compute here the growth of structure predicted by the HS model. The linear growth equation for modified gravity
 models is scale dependent and 
reads~\cite{Bean:2006up,Girones:2009nc}
\begin{equation}
\label{eq:growth}
\delta'' + \delta' \left( \frac 3 a + \frac{H'}{H} \right)
- \frac{3 \tilde{\Omega}a^{-3}}{\left(H/H_0\right)^2 (1+f_R)}
\frac{1-2Q}{2 - 3Q} \frac{\delta}{a^2} = 0~,
\end{equation}
where $' \equiv d/da$, $\delta$ is normalized such that $\delta \rightarrow a$ when $a \to 0$ and the factor $Q$ is given b
\begin{equation}
\label{eq:Q}
Q(k,a)= - 2 \; \left(\frac{k}{a}\right)^2 \frac{f_{RR}}{1+f_R}~.
\end{equation}

Note that in general relativity $Q$ is zero and therefore the linear
density growth is scale independent for all dark energy
models. However, for $f(R)$ models, the scale dependent $Q(k,a)$ 
induces a nontrivial scale dependence of the growth $\delta$.

\section{Current constraints}\label{sec:ii}

We exploit here current data from standard candles (SNIa) and standard rulers (BAO, CMB) in order to constrain the expansion history in the HS modified gravity model. Our analysis make use of the {\it Union2} survey \cite{Amanullah:2010vv}, which provides 557 SNIa useful to constrain luminosity distances. For the BAO, we exploit data from the Sloan Digital Sky Survey \cite{Eisenstein:2005su} and from 2-Degree Field (2dF) Galaxy Redshift Survey \cite{Percival:2009xn}. In addition, data from the WMAP7\cite{Komatsu:2010fb} satellite has been used to extract the CMB derived parameters $R$ (the shift parameter) and $l_A$ (parameter related to the first peak position). We build a global $\chi^2$ variable
\beq
\chi^2=\chi^2_{SN}+\chi^2_{BAO}+\chi^2_{CMB}~.
\enq

\begin{figure}[h!]
\begin{center}
\hspace*{-1cm}  
\includegraphics[width=8cm]{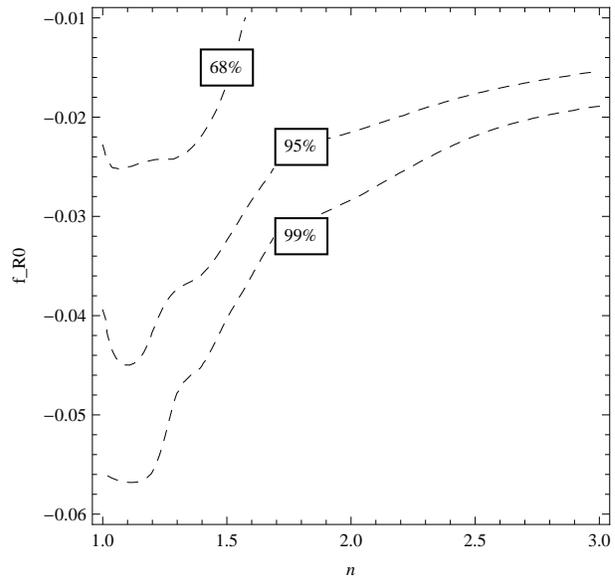}
\caption{Constraints on $n$ and $f_{R0}$ using current data.}
\label{fig:currRes}
\end{center} 
\end{figure}

\noindent The likelihood function is defined as
\begin{equation}
 L=e^{-\frac{\chi^2-\chi^2_{min}}{2}}~,
\end{equation}
and is marginalized over the present value for the Hubble expansion rate $H_0$.
The free parameters of the HS model are $n$ and $f_{R0}$. The value of the present dark matter energy density $\Omega_m$ has been fixed to $0.27$. Figure \ref{fig:currRes} shows the constraints on the HS model. Notice that current background data prefer small values of both $n$ and $f_{R0}$. These bounds are stronger than those presented in \cite{Martinelli:2009ek}, where $R$ and $l_A$ were not used and the SNIa and BAO datasets were from older catalogues.
In the following sections we show that future growth data will prefer higher value of $n$ values and the tension between future growth and background data will tighten the constraints on the HS model.
\section{Future data} \label{sec:iii}

\subsection{SN Luminosity distance data}
 
We shall exploit future SNIa data on the reduced magnitude
\begin{equation}
\mu=m-M=5\log_{10}d_L(z)+25~,
\end{equation}
being $d_L$ the luminosity distance given by Eq.~(\ref{eq:dl}). We consider here a mock catalog of 2,298 SNIa, with 300 SNIa uniformly distributed out to $z = 0.1$, as expected from ground-based low redshift samples, and an additional 1998 SNIa binned in 32 redshift bins in the range $0.1 < z < 1.7$, as expected from JDEM or similar future surveys~\cite{Kim:2003mq}. We have considered both intrinsic and systematic errors on the reduced magnitude.

\begin{figure}[h!]
\begin{center}
\hspace*{-1cm}  
\includegraphics[width=10cm]{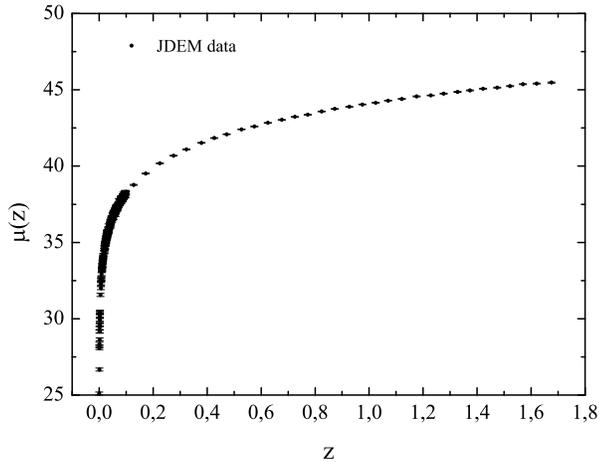}
\caption{Forecasted SNIa luminosity distance data for a JDEM-like survey.}
\label{fig:mockSN}
\end{center} 
\end{figure}

In the following, to produce mock SNIa luminosity distance and mock growth data from BAO surveys we shall assume a $\Lambda$CDM model with the following cosological parameters: $\Omega_{b}h^2=0.02258$, $\Omega_{c}h^2=0.1109$, $n_s=0.963$, $\tau=0.088$, $A_s=2.43\times10^{-9}$ and $\Theta=1.0388$, which correspond to the best-fit values from the WMAP seven year data analysis, see Ref.~\cite{Komatsu:2010fb}.

\subsection{Growth data}
Galaxy surveys measure the redshift of the galaxies, providing, therefore, 
the redshift space galaxy distributions. From those redshifts
the radial position of the galaxies are extracted. However, the inferred
galaxy distribution (and, consequently, the power spectrum) 
is distorted with respect to the true galaxy distribution, because 
in redshift space one neglects the peculiar velocities of the galaxies. 
These are the so called \emph{redshift space distortions}. 

In linear theory and with a local linear galaxy bias $b$ the 
relation between the true spectrum in real space and the spectrum 
in redshift space reads
\begin{equation}
\label{eq:kaiser}
P_\textrm{redshift}(\boldsymbol{k})=
\left(1 + \beta \mu_{\boldsymbol{k}}^2\right)^2 P(\boldsymbol{k}) \ , 
\end{equation}
where $\beta\equiv f/b$, being $f$ the logarithmic derivative of the linear growth factor $\delta(a)$ given by Eq.~(\ref{eq:growth}) and $\mu_{\boldsymbol{k}}$ is the cosine of the angle between the line of 
sight and the wavevector $\boldsymbol{k}$. Notice that perturbations with $\boldsymbol{k}$ perpendicular to the line of sight are not distorted.
The relation among real space and redshift space overdensities given
by Eq.~(\ref{eq:kaiser})  was first derived by Kaiser~\cite{Kaiser:1987qv} and it arises from the continuity equation, which relates the divergence of the peculiar velocity to the dark matter overdensity $\delta$. Redshift space distortions, then, relate peculiar velocities to the logarithmic derivative of the linear growth factor, $f$. A measurement of 
$\beta\equiv f/b$ will provide information on the growth of structure formation if the galaxy bias $b$ is known. 

In our analysis we have computed the forecasted errors for $f$ assuming a $\Lambda$CDM fiducial model by means of a Fisher matrix analysis, marginalizing over the bias $b$. 
 We focus here on the BAO experiments BOSS~\cite{Eisenstein:2011sa} and Euclid \cite{Refregier:2006vt}. For the BOSS (Euclid) experiment we assume six (nineteen) redshift bins ranging from $z=0.15$ to $z=0.65$ ($z=0.15$ to $z=1.95$) and a galaxy survey area of 10000 (20000) deg$^2$. The mean galaxy densities for these two experiments are considered to be constant with values of $2.66 \times 10^{-4}$ and $1.56 \times 10^{-3}$~$h$~Mpc$^{-3}$ for the BOSS and the Euclid surveys respectively. 

\begin{figure}[h!]
\begin{center}
\hspace*{-1cm}  
\includegraphics[width=10cm]{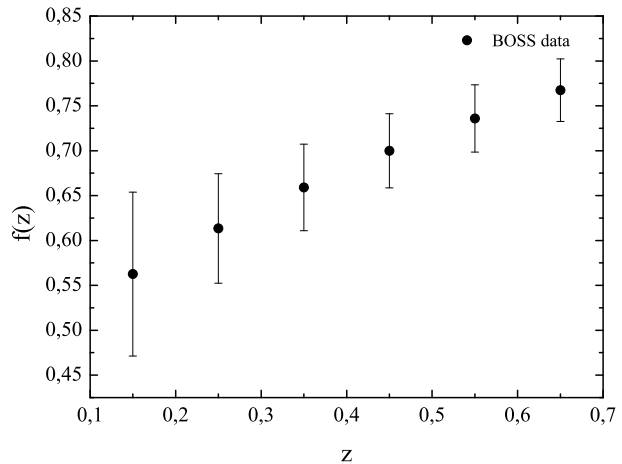}
\caption{Forecasted BOSS data for $f$, the logarithmic derivative of the linear growth factor, as a function of the redshift.}
\label{fig:boss_growth}
\end{center} 
\end{figure}

\begin{figure}[h!]
\begin{center}
\hspace*{-1cm}  
\includegraphics[width=10cm]{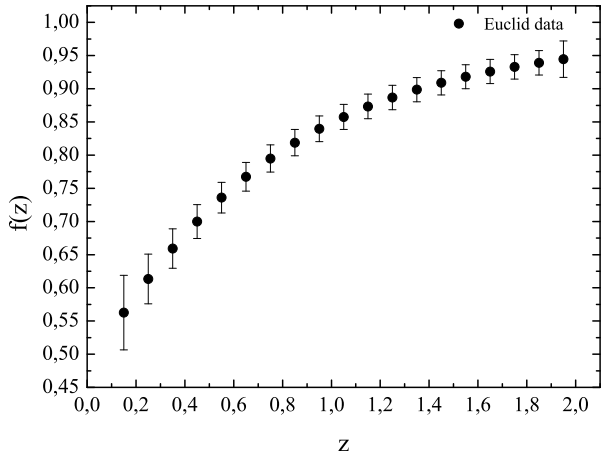}
\caption{Forecasted Euclid data for $f$, the logarithmic derivative of the linear growth factor, as a function of the redshift.}
\label{fig:euclid_growth}
\end{center} 
\end{figure}

We have combined the galaxy survey fisher matrices with the CMB Planck~\cite{:2006uk} Fisher matrix, see Ref~\cite{Verde:2005ff}. The expected errors on 
$f$ are depicted in Figs.~\ref{fig:boss_growth} and \ref{fig:euclid_growth} and are similar to the ones obtained using the numerical tools of Ref.~\cite{{White:2008jy}}.

\section{Future Data Analysis and Results} \label{sec:iv}

In this section we perform a numerical fit to the HS model using the mock data sets described above by means of a $\chi^2$ analysis. We compute two different 
$\chi^2$ functions, one associated to growth of structure ($\chi^2_{growth}$) and the other related to SNIa luminosity distance data, $\chi^2_{SN}$.  The final likelihood will exclusively depend on the free parameter of 
the HS model, $n$ and $f_{R_0}$, thus for the two different $\chi^2$ analyses ($\chi^2_{growth}$ and $\chi^2_{SN}$) and well as for the combined one 
($\chi^2_{tot}=\chi^2_{growth}+\chi^2_{SN}$) we marginalize over $\tilde{\Omega}_m$. For the growth of structure we compute the $\chi^2_{growth}$ at different scales $k$ scales and perform a mean over $k$ from $k=0.01h$~Mpc$^{-1}$ to $k=0.1h$~Mpc$^{-1}$. 
\begin{figure}[h!]
\begin{center}
\includegraphics[width=8cm]{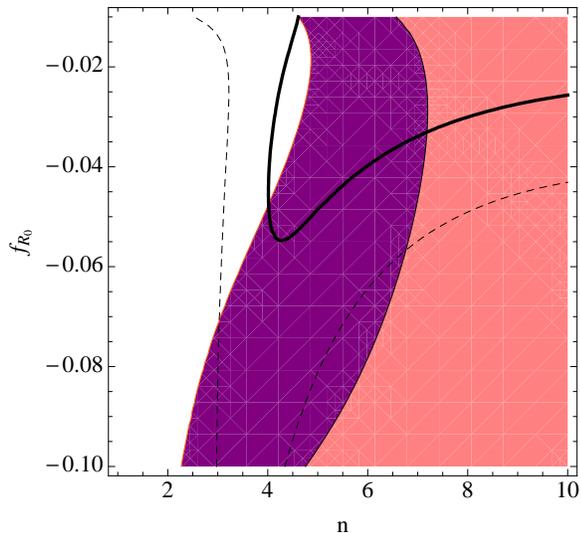}
\caption{1 and 2-$\sigma$ contours in the $(n,f_{R_0})$ plane arising from SNIa measurements of the luminosity distance (black solid and dashed lines) and those arising from the BOSS measurements of the linear growth of structure (filled regions).}
\label{fig:boss}
\end{center}
\end{figure}

\begin{figure}[h!]
\begin{center}
\includegraphics[width=8cm]{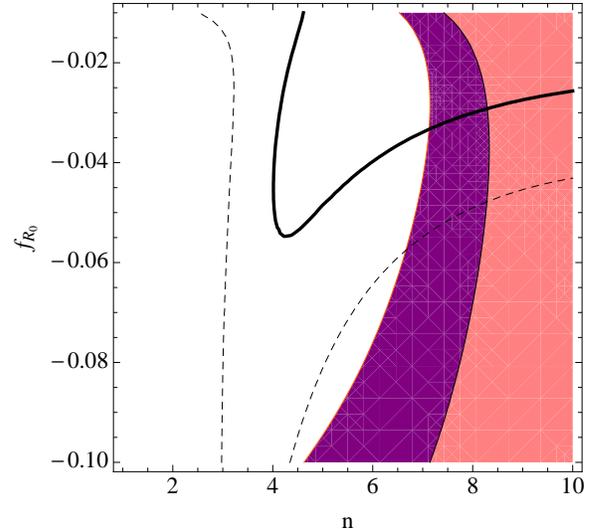}
\caption{1 and 2-$\sigma$ contours in the $(n,f_{R_0})$ plane arising from SNIa measurements of the luminosity distance (black solid and dashed lines) and those arising from the Euclid measurements of the linear growth of structure (filled regions).}
\label{fig:euclid}
\end{center}
\end{figure}

Figure \ref{fig:boss} shows the 1 and 2-$\sigma$ contours on the $(n,f_{R_0})$ plane arising from the analysis of SNIa luminosity distance data as well as from measurements of the linear growth of structure from the BOSS experiment.  
The empty regions depicted by the dotted and solid lines show the results from SNIa luminosity distance data, while the filled regions depict the results from the measurements of the linear growth of structure. Notice that background and growth measurements are complementary, allowing different regions in the parameter space. The combination of both data sets provide a very powerful tool to severely constrain the HS model. 

\begin{figure}[h!]
\begin{center}
\begin{tabular}{cc}
\includegraphics[width=8cm]{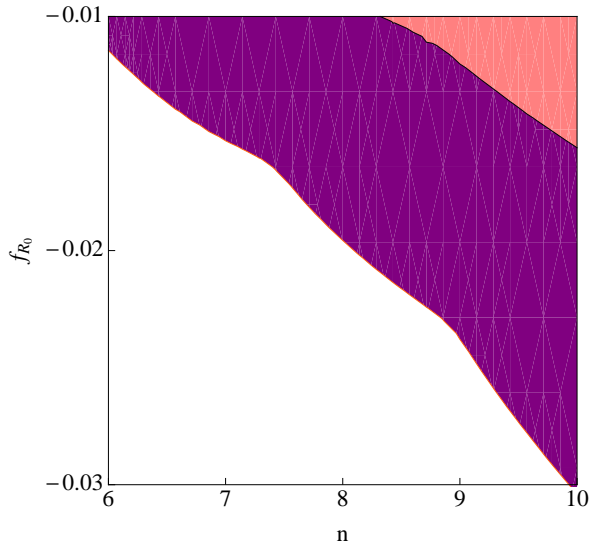}
\end{tabular}
\caption{1 and 2-$\sigma$ contours in the $(n,f_{R_0})$ plane arising from BOSS  measurements of the linear growth of structure combined with SNIa 
luminosity distance data from the JDEM survey. Note that the axis ranges are different to those of Fig.~\ref{fig:currRes}.}
\label{fig:boss_comp}

\end{center}
\end{figure}

\begin{figure}[h!]
\begin{center}
\begin{tabular}{cc}
\includegraphics[width=8cm]{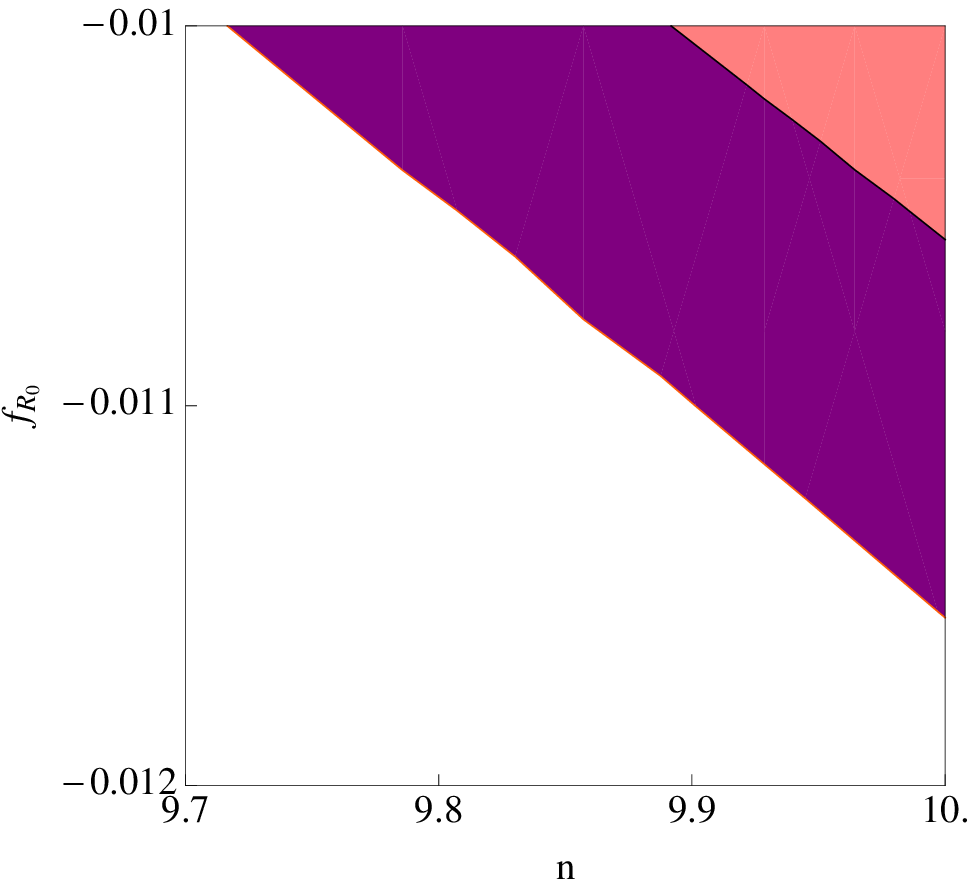}
\end{tabular}
\caption{1 and 2-$\sigma$ contours in the $(n,f_{R_0})$ plane arising from Euclid  measurements of the linear growth of structure combined with SNIa 
luminosity distance data from the JDEM survey. Note that the axis ranges are different to those of Figs.~\ref{fig:currRes} and \ref{fig:boss_comp}.}
\label{fig:euclid_comp}
\end{center}
\end{figure}

\begin{figure}
\begin{center}
\begin{tabular}{cc}
\includegraphics[width=8cm]{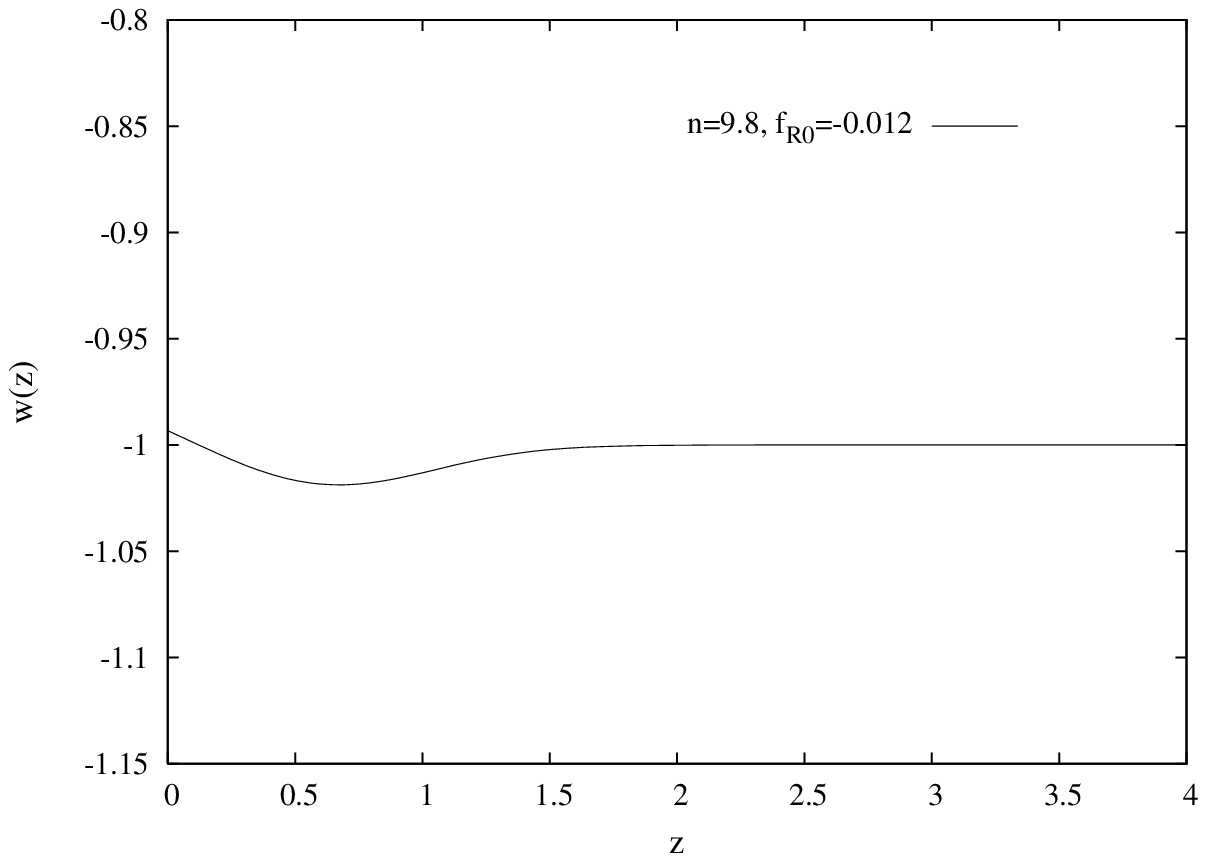}\\
\includegraphics[width=8cm]{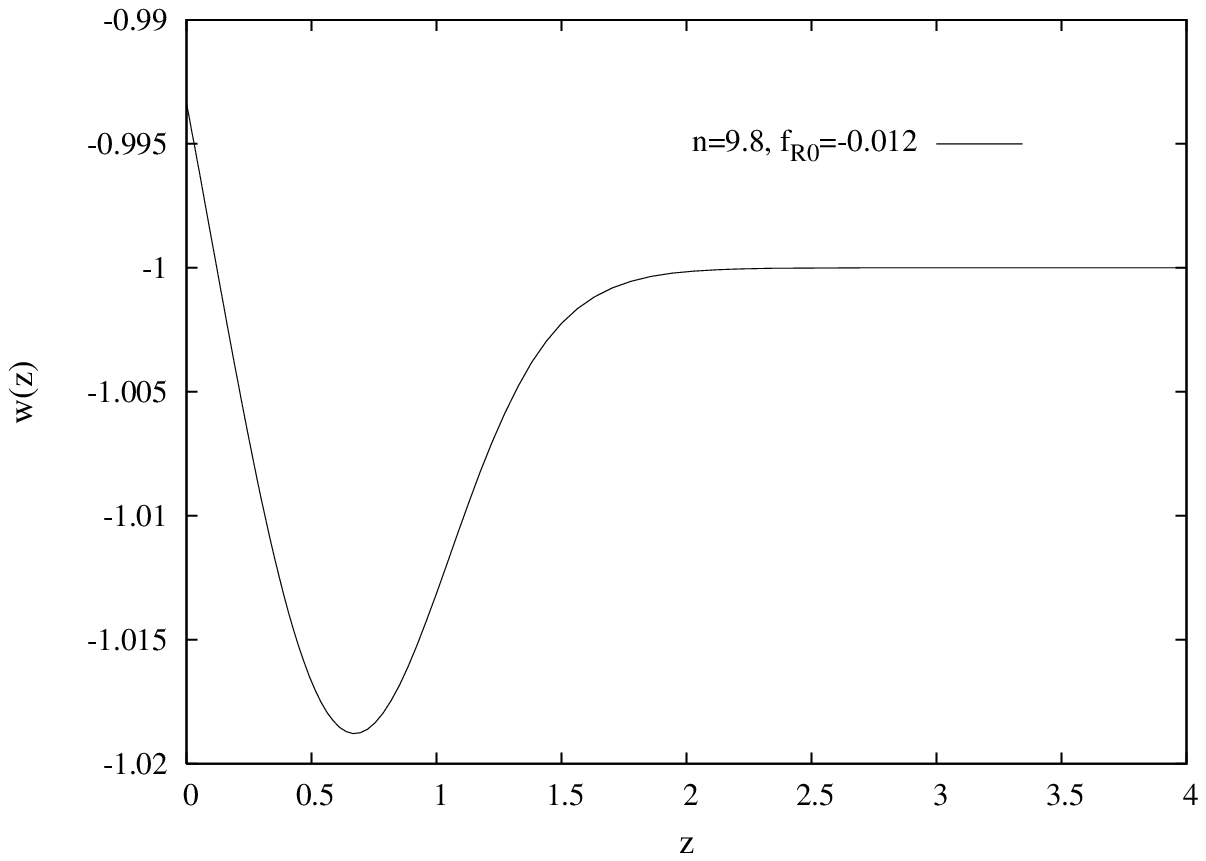}
\end{tabular}
\caption{The top and bottom panels depict the effective equation of state $w(z)$ for $n=9.8$ and $f_{R_0}=-0.012$, values that are in agreement with Euclid growth data. The top panel shows $w(z)$ using the same ranges than those use in  Figs.~\ref{wn1} and \ref{wn2}. The bottom panel shows an axes range that allows to appreciate the behaviour of $w(z)$.}
\label{fig:wok}
\end{center}
\end{figure}

Figure \ref{fig:euclid} shows the equivalent to Fig.~\ref{fig:boss} but considering measurements of the linear growth of structure from the Euclid galaxy survey. Notice that the currently allowed region will be ruled out if the true cosmology is a $\Lambda$CDM scenario. Therefore, the BOSS and Euclid surveys, combined with background SNIa luminosity distance measurements, will be able to exclude a very large region of the parameter space of the HS model, allowing only for very high values of $n$ for which $c_1/c^2_2\rightarrow 0$, see Figs.~\ref{fig:boss_comp} and \ref{fig:euclid_comp} in which the combined analyses $\chi^2_{tot}$ are shown.  Figure \ref{fig:wok} illustrates that, in the limit of $c_1/c^2_2\rightarrow 0$, i.e. for very large values of $n$, $f(R)$ is equivalent to a cosmological constant. Therefore, future data from the Euclid galaxy survey combined with SNIa measurements will be able to recover the true fiducial cosmology even if the data is fitted to a non constant $f(R)$. 

\section{Conclusions} \label{sec:v}
The Hu-Sawicki modified gravity scenario is analyzed with current background expansion data, updating previous results \cite{Martinelli:2009ek}. Our analysis shows that background data allow for a region of the parameter space where the parameter $n$ of the model can have small values, inducing a late time dynamical behavior on the effective dark energy equation of state $w(z)$.  Future constraints on the Hu-Sawicki modified gravity model arising from measurements of the linear growth of structure and of SNIa luminosity distances are also presented.  While luminosity distance data allow for small values of $n$, the growth of structure data prefer higher values of $n$. The combination of these two observables allows to tightly constrain the HS model. We have generated mock growth and luminosity distance data data for a fiducial $\Lambda$CDM model and fitted these data in a Hu-Sawicki modified gravity scenario. The reconstructed effective dark energy equation of state almost is identical to that of a $\Lambda$CDM model. 

\section*{Acknowledgments}
 O.M. is supported by AYA2008-03531 and the Consolider Ingenio-2010 project CSD2007-00060.  This work is supported by PRIN-INAF, ”Astronomy probes fundamental physics”. Support was given by the Italian Space Agency through the ASI contracts Euclid-IC (I/031/10/0).

\bibliography{bibfr}

\end{document}